# Low temperature laser scanning microscopy of a superconducting radio-frequency cavity

G. Ciovati[1,a)], Steven M. Anlage[2], C. Baldwin[1], G. Cheng[1], R. Flood[1], K. Jordan[1], P. Kneisel[1], M. Morrone[1], G. Nemes[3], L. Turlington[1], H. Wang[1], K. Wilson[1], and S. Zhang[1]

[1]*Thomas Jefferson National Accelerator Facility, 12000 Jefferson Avenue, Newport News, Virginia 23606, USA*

[2]*Center for Nanophysics and Advanced Materials, Department of Physics, University of Maryland, College Park, Maryland 20742-4111, USA*

[3]*ASTiGMAT^{TM}, 3409 Pecky Cedar Ct., Sacramento, CA 95827, USA*

An apparatus was developed to obtain, for the first time, 2D maps of the surface resistance of the inner surface of an operating superconducting radio-frequency niobium cavity by a low-temperature laser scanning microscopy technique. This allows identifying non-uniformities of the surface resistance with a spatial resolution of about one order of magnitude better than with earlier methods and surface resistance resolution of ~ 1 $\mu\Omega$ at 3.3 GHz. A signal-to-noise ratio of about 10 dB was obtained with 240 mW laser power and 1 Hz modulation frequency. The various components of the apparatus, the experimental procedure and results are discussed in detail in this contribution.

---

a) Author to whom correspondence should be addressed. Electronic mail: gciovati@jlab.org.

# I. INTRODUCTION

Superconducting radio-frequency (SRF) cavities have been used in particle accelerators for more than thirty years and their performance, in terms of achievable accelerating gradients, has been continuously improving over the years. Niobium, either as a thin film or "bulk" material, has been the material of choice to produce SRF cavities because of its highest critical temperature, $T_c$, among all elemental superconductors and its highest value of the lower critical magnetic field, $H_{c1}$.

The most powerful diagnostic technique which has been applied so far to study anomalies in the surface resistance of SRF cavities at cryogenic temperatures has been temperature mapping. This consists of covering the outer cavity surface with an array of thermometers, typically custom built sensors based on carbon resistors, to measure the local temperature while increasing the RF power in the cavity [1]. These measurements provide an image of the heat due to RF losses generated on the inner cavity surface. The most modern implementation of the thermometry technique was done at Cornell University [2]. The spatial resolution of the technique is related to the density and size of the thermometers and is of the order of 1 cm. Typical temperature maps of SRF cavities show a significant non-uniformity in the heat dissipation, which is referred to as "hot-spots" [3], usually located in the regions of high surface magnetic field. The origin of such hot-spots is not yet clear and is the subject of ongoing research at several laboratories and universities throughout the world. Hot-spots are areas of the surface where premature thermal quench of the superconducting state occurs [4] or the excessive heating causes a reduction of the quality factor, $Q_0$, of the resonator (a phenomenon referred to as "Q-slope") [5]. In either case, the presence of hot-spots limits the operational accelerating gradient of SRF cavities.

Low temperature laser scanning microscopy (LTLSM) is a well-known technique to characterize loss mechanisms in high-temperature superconductors (HTS) [6]. The technique consists, essentially, of a point-by-point raster scanning of a focused laser beam onto the surface of the sample under study, then measuring the photo-response PR(x, y), induced by the interaction of the light with the superconductor, as a function of the laser spot position (x, y). Several different techniques can be used to convert the PR(x, y) signal to obtain the local LSM contrast voltage δV(x, y), which allows building a 2D image of the optical, RF, electronic and other properties of the superconductor. The probe intensity is modulated in amplitude with the frequency $f_M$ and lock-in detected to enhance the signal-to-noise ratio and hence the contrast. A review of this technique can be found in Ref. [6].

In this contribution we describe an experimental setup which allows, for the first time to our knowledge, obtaining 2D images of the local surface resistance in regions of an SRF cavity with millimeter-size resolution using the LTLSM technique. A detailed description of the various optical, RF and mechanical components of the system is given in Sec. II. The experimental procedure is described in Sec. III. Some experimental results are highlighted in Sec. IV, while discussion and conclusions are given in Sec. V and VI respectively.

# II. EXPERIMENTAL APPARATUS

The SRF cavity is attached to a test stand which is inserted in a vertical cryostat, 71 cm in diameter and 275 cm deep into the ground. A 532 nm laser is placed on the test stand top plate. The laser beam is directed into a vacuum pipe inside the cryostat, reflected off two rotatable, scanning mirrors placed

~154 cm below the top plate, then contacts the surface of the SRF cavity flat plate. Optical components on the top plate allow adjustment of both the beam power and its diameter. The two scanning mirrors are centrally mounted on the shafts of two stepper motors inside a vacuum chamber. Figure 1 shows a schematic and a picture of the experimental apparatus and the various components are described in details that follow.

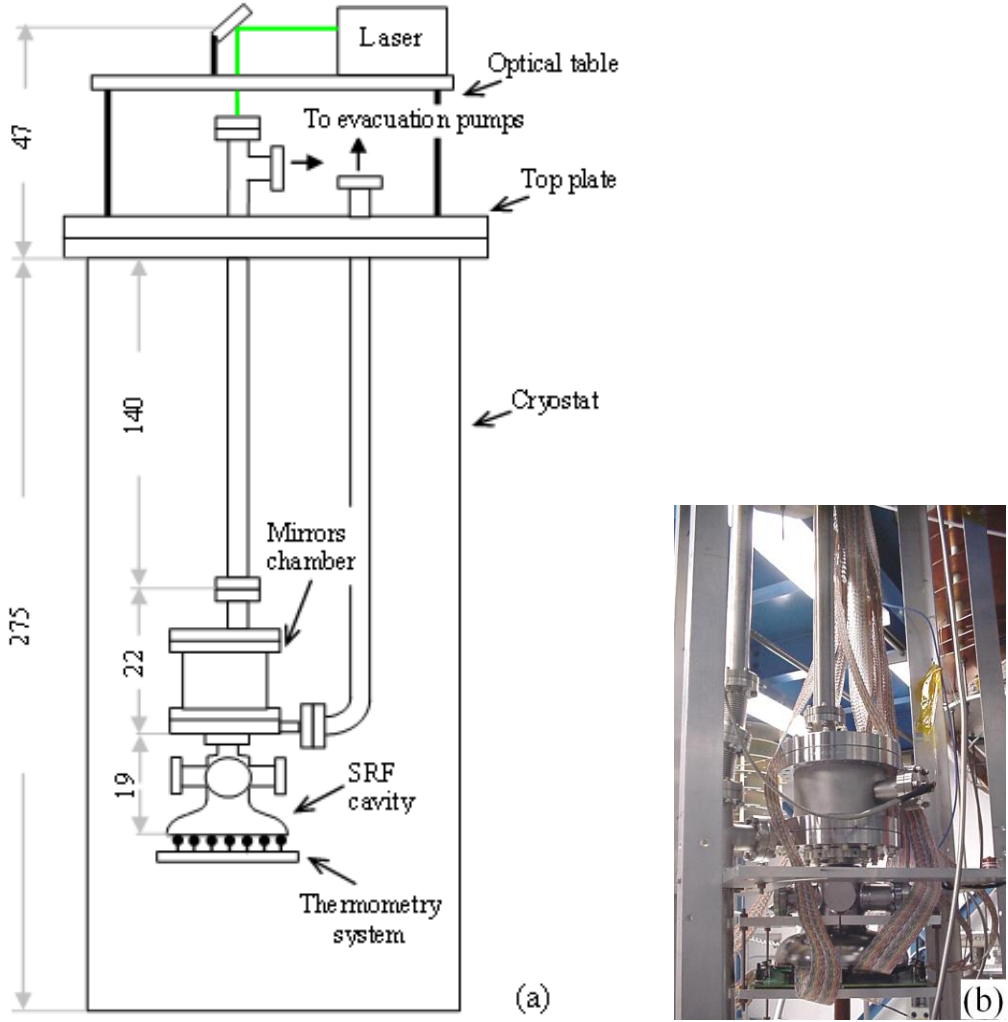

FIG. 1. (Color online) (a) Schematic of the experimental apparatus. Dimensions are in cm. (b) Picture of the mirrors chamber, SRF cavity and thermometry system assembled on the vertical test stand.

## A. SRF cavity design and fabrication

The SRF cavity consists of a half-cell of the TESLA shape [7] with a flat plate welded at the equator. The equator diameter is 210.2 mm and the iris diameter is 70 mm. This cavity geometry had been used to fabricate prototype cavities for the development of SRF photoinjectors [8]. The resonant frequency of the accelerating mode $TM_{010}$ is 1.3 GHz, however a different resonant mode was used for this setup. The TE$_{011}$ mode at 3.3 GHz has the advantages of having, ideally, zero surface electric field, thereby avoiding field emission to limit the cavity performance, while having high surface magnetic field in a region of the surface which can be accessed by the laser beam. Coupling of the RF power into the cavity is done with an "L-shaped" copper antenna inserted into a 40 mm diameter side port. A small sample of the cavity stored energy is coupled out through a shorter copper antenna inserted into a 10 mm diameter side port, located 90° and coplanar to the input side-port. The length of the cut-off tube was minimized to allow the laser beam to reach as far as possible from the center of the flat plate while the diameter of the cut-off tube had to be reduced from 70 mm to 40 mm near the end to reduce the amplitude of the RF field at the end of the cut-off tube. The ratio of the RF power dissipated on the cavity surface to the power dissipated on "external" components, such as blank-off discs and antennae is called $Q_{ext}$ and the length of the cut-off tube and side ports was designed to have $Q_{ext} > 1 \times 10^{12}$ for all stainless steel blank-off discs. The first prototype cavity was built from 3.125 mm thick large-grain niobium from CBMM, Brazil. The residual resistivity ratio is ~200. The flanges were made of Ti-45Nb and different vacuum seal configurations were used: Cu gasket and knife-edge Conflat design for the flange of the 40 mm diameter side port, AlMg$_3$ gasket for the 10 mm side port flange and In wire for the flange on the cut-off tube. All the cavity parts were joined by electron beam welding. Two niobium stiffening ribs, 25.4 mm tall, 3 mm thick had to be welded across the diameter face of the first cavity flat plate, joined at 90° at the center, to avoid face distortion when the inside volume of the cavity is evacuated.

The surface preparation of the cavity consisted of ~100 μm material removal from the inner surface by buffered chemical polishing (BCP) and heat treatment in an ultra-high vacuum (UHV) furnace at 800 °C for 3 h. Cryogenic RF tests of the cavity at 2.0 K revealed the presence of multipacting, limiting the cavity peak surface magnetic induction, $B_p$, to 40-55 mT. 3D multipacting simulations confirmed the presence of conditions for multipacting at field levels consistent with the experimental data. The analysis showed that the 40 mm diameter side-port introduces a very small perturbation of the ideal electromagnetic field configuration of the TE$_{011}$ mode such that a finite surface electric field component normal to the surface is present on the cavity flat plate and can accelerate electrons emitted during the multipacting process. The problem could be mitigated by adding 4 side-ports of the same 40 mm diameter, welded at 90° and coplanar to one another to reduce the perturbation of the cylindrical symmetry of the TE$_{011}$ mode. Simulations indicate that multipacting conditions for this new configuration are possible at $B_p$-values of about 100 mT. Details about the experimental results and simulation study can be found in [9]. Based on the experience gained with the first prototype, a second cavity from the same CBMM niobium ingot was built with modified cut-off tube and side-ports and the completed cavity is shown in Fig. 2. The surface resistance maps were taken on regions of the flat plate of this cavity. The cavity flat plate with stiffening ribs was machined from a solid piece of Nb ingot. All flanges are made of Ti-45Nb and In wire is used as gasket material. Figure 3 shows the intensity of the magnetic field on the cavity surface for a stored energy of 1 J as calculated with the 3D electromagnetic field solver Microwave Studio [10]. Cryogenic RF tests at 2.0 K of this cavity did show a multipacting barrier at 90-100 mT but it was possible to process it up to $B_p \sim$ 140 mT on one occasion.

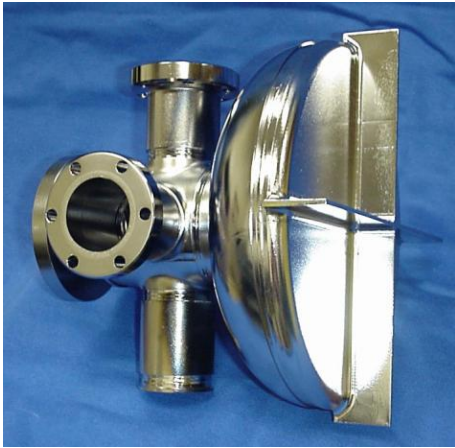

FIG. 2. (Color online) Picture of the SRF niobium cavity.

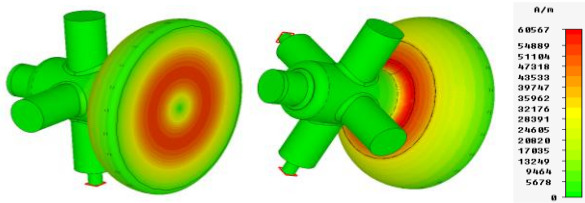

FIG. 3. (Color online) Calculated distribution of the surface magnetic field in the $TE_{011}$ mode at 3.3 GHz for a cavity stored energy of 1 J.

### B. Thermometry system

A thermometry system to measure the temperature of the cavity flat plate during high-power RF testing at 2 K was designed and built. The thermometry system allows identifying the RF hotspots which will be analyzed using the LTLSM technique as well as locating the position of the laser beam on the cavity flat plate and verifying the movement of the beam during laser scanning. The system consists of 128 carbon resistors (Allen-Bradley 100 Ω, 1/8W), each encapsulated in a G-10 housing and mounted on pogo-sticks [2] inserted in an aluminum plate, and through a printed circuit board (PCB). Apiezon N grease is used to improve the thermal contact between the thermometers and the outer cavity surface. The sensors are evenly distributed along seven concentric "rings" such that the distance between each sensor and its neighbors is 12-20 mm. The sensors' leads are soldered to copper traces on the PCB board which route the signals to eight 37-pin IDC connectors on the side of the board. Twisted pair flat ribbon cables inside the cryostat connect to the PCB board on one end and to two feed-through cans (with four connectors each) mounted on the top plate on the other end. Eight shielded jacketed ribbon cables bring the signals from the feed-through cans to four custom PCBs which interface to four 32-channel analog input modules (National Instruments, SCXI-1100). The signals are digitized by a 16-bit analog to digital converter (National Instruments, PCI-6250 M) inside a computer. The temperature sensors are connected in parallel to a voltage source (Keithley 2400 source-meter) which provides an excitation current of 4 μA to each sensor. The typical sensor resistance at 2.0 K is ~ 6-7 kΩ with sensitivity dR/dT(2 K) ~ 10 Ω/mK. The data acquisition time for a single scan of the 128 sensors is ~10 ms. Data acquisition is accomplished with LabVIEW programs. Figures 4(a)-(b) show a picture of the PCB board with all the sensors and of the feed-through cans.

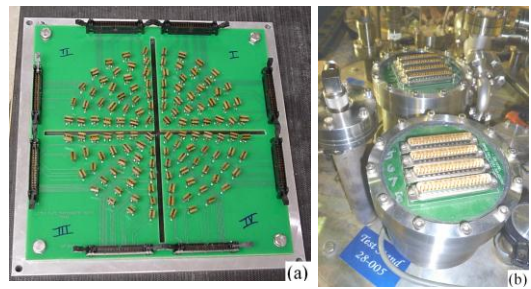

FIG. 4. (Color online) Pictures of the PCB thermometry board with 128 thermometers (a) and of the feed-through cans on the test stand top plate (b). The thermometer array is attached to the flat circular surface of the cavity shown in Fig. 2.

### C. Mirrors' chamber

The vacuum chamber which contains two remotely-rotatable scanning mirrors, each mounted to a high-vacuum mini-stepper motor, was designed and built. The chamber consists of a stainless steel tube, 127 mm tall, 152.4 mm inner diameter (ID), with two Conflat flanges of 203 mm outer diameter (OD) welded onto each end. Two ports, 38 mm OD, are welded to the side of the chamber and each has a 70 mm OD stainless steel Conflat flange welded onto the end. A two-pair K-type thermocouple feed-through (Solid Sealing Technology, Part No. FA13091) is bolted to one of the side-ports' flange, while a 2×9-pin feed-through (Accu-Glass Products, Part No. 100012) is bolted to the other side-ports' flange. Scanning of the laser beam in the x-y directions is accomplished with two flat mirrors placed in aluminum holders axially mounted to the shafts of two stepper motors (Phytron, Part No. VSS 19.200.0.6-NSSN-KTC-HV). The two stepper motors (19 mm diameter, 26.5 mm long) are mounted rotated 90° with respect to each other. The laser beam is reflected off the first mirror (20 mm × 20 mm × 3 mm in size) whose plane is at a 45° angle from the vertical beam direction and is then directed sideward toward the second mirror (40 mm × 20 mm × 3 mm in size), which is also at a 45° angle with respect to the vertical axis. At the second mirror the beam is reflected downwards toward the Nb cavity. The center of the second mirror is at the same height as the center of the first mirror but is horizontally shifted by 25 mm. A schematic of the mirrors' chamber is shown in Fig. 5. Rotation of the first mirror allows scanning in the x-direction, while rotation of the second mirror allows scanning in the y-direction.

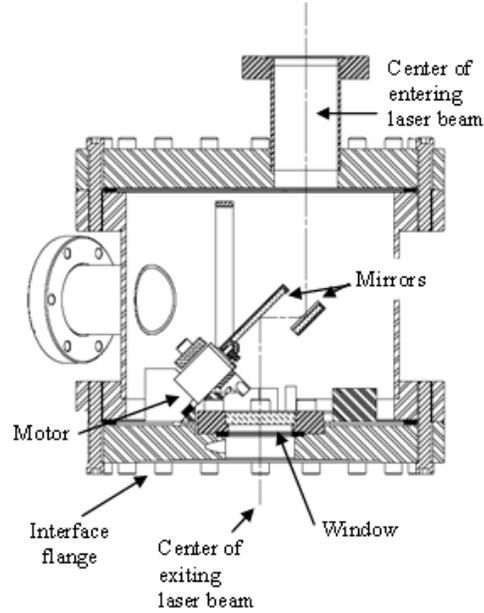

FIG. 5. Schematic of the mirrors' chamber.

The mirrors, made of fused silica, were fabricated by Casix, China, and their surface was coated with a high reflectance coating at 532 nm for 45° angle of incidence (AOI), unpolarized beam. The motors have parallel windings with 0.6 A/phase, 200 steps/revolution, a type K thermocouple attached to the windings and a low stray magnetic field rotor configuration. In order to reduce the cost of the motors and for better reliability it was decided not to use cryogenically rated motors but to use "standard" high-vacuum motors.

Because the outer surface of the mirrors' chamber is immersed in the liquid helium bath at 2.0 K, a proper support structure had to be designed to maintain the motors' temperature above the lowest rated value of -20 °C. A Kapton flexible heater (Watlow, 5 W, 25.4 mm × 25.4 mm in size) is taped to the body of each motor with Kapton tape. Each motor with heater is mounted to an aluminum frame which is then mounted to a G-10 ring. The G-10 ring has three slots machined on the bottom surface which allow mounting the whole assembly onto steel dowel-pins pressed into the inner face of the chambers' lower flange. This design allows minimizing heat conduction by using G-10 for the base, minimizing the mounting contact area with the chamber's walls, while minimizing irradiative heating from the motors' heaters to the chamber's walls by using internally-polished aluminum clamps. A picture of the G-10 support structure with the motors, mirrors and heaters in place (minus clamp-brackets) is shown in Fig. 6.

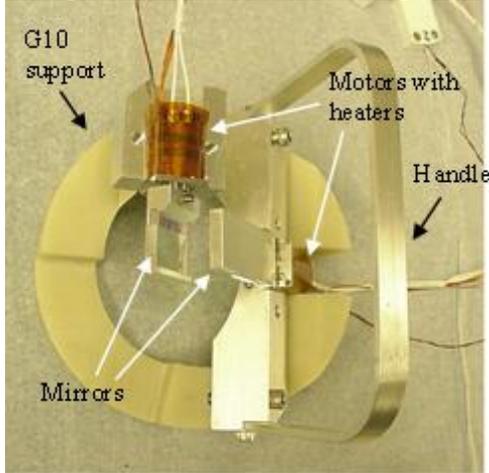

FIG. 6. (Color online) Picture of the motors' support structure.

After cooldown of the mirrors' chamber to 2 K, it was determined that a heater power of ~ 0.5-1.25 W was sufficient to maintain the motors' temperature within the range 5 – 40 °C. The same type of feed-throughs bolted to the chambers' side-ports were used on the test stand top plate to bring the signals from motors' temperatures, windings and heaters' leads from inside the cryostat to the air side. The motors are controlled by a 2-axis motion controller (National Instruments, PCI-7340) inside a PC and power supply (National Instruments, MID-7602). The motors' temperatures are measured with two thermocouple measurement devices (National Instruments, USB-TC01) plugged into a PC. A programmable DC power supply (Instek, PSS-3203, 32V/3A) is used to provide current to the heaters, which are connected in parallel. A LabVIEW program is used to control the motors' (mirrors') positions and the heaters' power supply.

The largest radius that the laser beam can reach from the center of the cavity flat plate is about 40 mm, close to the center of the "ring" with high surface magnetic field shown in Fig. 3, at a radius of 44 mm. A 40 mm radial position of the laser beam corresponds to a rotation of about ± 4° of the mirrors from their initial position. The motors power supply allows up to 256 microsteps per one full motor step, so that the x-y spatial resolution of the beam position on the cavity flat plate is 69 μm and 44 μm, respectively.

The interface flange between the mirrors' chamber and the SRF cavity is a 203 mm OD Conflat blank-flange with a thru-hole and knife edge configured for a 70 mm OD Conflat flange being machined concentrically on the same face which has the larger diameter knife-edge. A standard 35.6 mm diameter viewport made of Corning 7056 glass, with an antireflective (AR) coating at 532 nm, 0° AOI, mounted on a 70 mm OD Conflat flange (Accu-Glass Products, Part No. 112402) is bolted to the interface flange. The viewport allows isolating the cavity volume from the volume of the mirrors' chamber, in order to maintain the cleanest conditions on the cavity surface, which are needed to obtain a high $Q_0$ value. During initial RF tests it was found that the viewport had a stray magnetic field from the nickel-iron sleeve used for the stainless steel to glass seal transition, which degraded the cavity $Q_0$ by one order of magnitude. A new viewport was designed which consists of a 38.1 mm diameter UV-grade fused silica laser window, 3.175 mm thick and AR coated at 532 nm, 0° AOI (CVI Melles Griot, Part No. W2-PW-1508-UV-532-0), which is compressed between a stainless steel retainer ring and a 70 mm OD Conflat flange. The Conflat flange has a seal-recess and a 33 mm diameter hole in the center and 1 mm diameter In wire is used to seal the window to the stainless steel retainer ring and flange. A picture of the viewport bolted to the interface flange is shown in Fig. 7. A 9.5 mm diameter hole is drilled on one side through the thickness of the interface flange to the machined hole in its center, to provide a flanged port for the evacuation of the SRF cavity. The bottom surface of the interface flange has a bolt-hole pattern and seal-recess to mount the flange to the cut-off tube of the cavity. 1.5 mm diameter In wire is used to seal the cavity flange to the interface flange.

The top of the mirrors' chamber is bolted to a 203 mm OD Conflat blank-flange which has a counter-bored access hole welded to a 25 mm diameter half-nipple with a 70 mm OD Conflat flange welded to its end. A 140 cm long vacuum tube connects the mirrors' chamber top flange to the bottom face of the test stand top plate.

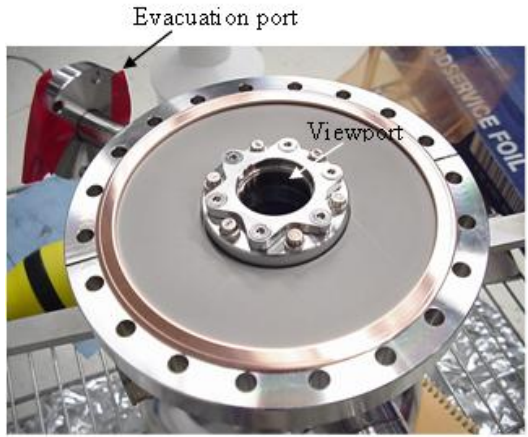

FIG. 7. (Color online) Picture of the interface flange with viewport at the center.

## D. Optics design

A 532 nm diode-pumped solid-state laser with 10 W output power and 5% power stability (Laserglow, part No. LRS-0532-PFW-10000-05) was used for the experiments. The laser power can be modulated by an auxiliary TTL signal up to 30 kHz. The absorptivity (or absorbance) of niobium was measured to be 40-50% at 4.2 K and 15° AOI at 532 nm [11]. The output aperture of the high-power laser is inside an "attenuator box" which allows obtaining a laser beam average power between ~ 3 mW and 9.8 W at its output. This is done by placing a $\lambda/2$ waveplate on a rotating stage after the laser, followed by a polarizing beamsplitter cube: the p-polarization is transmitted straight through the cube and is absorbed by a beam dump, while the s-polarization is reflected at 90°. An optical filter with neutral density (ND) = 2 is used to further attenuate the 90° beam to the lowest power level. By rotating the $\lambda/2$ waveplate between 0° and 46° it is possible to adjust the beam power output between its maximum and its minimum value, respectively. A proximity switch is attached to the waveplate rotating stage providing an interlock signal which allows laser operation only when the waveplate position is at 46°. By using such attenuator box it is possible to perform the alignment of the laser beam with minimum protective measures for the eye. It also provides the flexibility of adjusting the laser power during the experiments while maintaining a constant laser beam shape and size at different power levels on the target. A schematic of the optics on the test stand top plate and a picture of the setup are shown in Figs. 8(a)-(b).

The optical beam spatial parameters were obtained by measuring the beam profile at different distances from the laser output, using a CCD camera (COHU 4812) with beam analyzer software (Spiricon, LBA-PC v.4.25). The results indicate general astigmatic-type laser beam, which, for simplicity, can be approximated to a stigmatic-type one [12]. The approximately stigmatic laser beam has the parameters: waist position ($z_0$) at the exit aperture of the laser, waist diameter ($D_0$) of 3.9 mm, divergence (full angle, $\theta$) of 1.4 mrad, a Rayleigh range ($z_R$) of 2.8 m and a beam propagation ratio ($M^2$) of 8 (quantities and notations according to ISO 11146-1,2,3 [13]). Based on these beam parameters, an optical system was designed to achieve a beam diameter on target, $D_L$, defined at $1/e^2$, adjustable between ~0.7 mm and ~4 mm. The optical components have to be mounted on an optical table mounted on the test stand top plate, along with the laser and the attenuator box. The approximate available length along the horizontal direction, from the output of the laser box to the 90° mirror (shown in the schematic in Fig. 1) which directs the beam downward towards the mirrors' chamber, is 382 mm, while the available length along the vertical direction is ~ 2304 mm. The adjustment of the beam diameter on target is achieved by placing a diverging lens with -76.4 mm focal length (CVI Melles Griot, part No. LUK-25.0-35.2-UV-425-675) on a translation stage, close to the output of the attenuator box, followed by a converging lens with 305.8 mm focal length (CVI Melles Griot, part No. LUP-50.0-140.9-UV-425-675), and moving longitudinally the diverging lens. The distances between the lenses, the 90° mirror and the output of the laser box were calculated using optical geometric analysis and full beam propagation analysis and both gave similar results. After mounting the components on the optical table at distances as close as possible to the calculated values, the laser beam was centered to the 140 cm long pipe below the top plate. Then the mirrors' chamber was attached and the beam profile was measured with the CCD camera at the cavity plate position, below the mirrors' chamber, for different values of the distance between the 2 lenses. The smallest beam diameter of 0.87 mm corresponds to a distance of ≈259.1 mm between the flat surfaces of the lenses (7.62 mm travel of the translation stage), while the largest obtained diameter of 3.0 mm corresponds to a distance of ≈266.7 mm between the flat surfaces of the lenses (zero travel of the translation stage).

The challenge of aligning the beam with the geometrical axis of the 140 cm long vacuum tube was met by placing a glass retroreflector (also called "corner cube") in a holder, attached at the bottom of the tube, which centers its vertex with respect to the tube's axis. The beam position is adjusted by moving the gimbaled mounts which hold the two 90° mirrors on the optical table on the top plate (a second 90° mirror was placed between the two lenses to maintain the beam path within the perimeter of the top plate) such that the beam reflected backward from the retroreflector at the bottom of the tube overlaps with the forward beam.

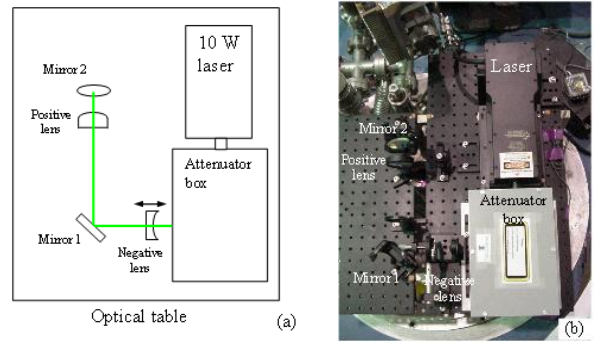

FIG. 8. (Color online) Schematic (a) and top-view picture (b) of the optical components on the test stand top plate.

## III. EXPERIMENTAL PROCEDURE

Each individual sensor of the thermometry system is calibrated in the temperature range 3.5 – 2 K. Two calibrated germanium resistors (Lakeshore, model GR-200A-250) immersed in the liquid helium bath are used as reference. The cavity is locked at the resonant frequency with a phase-locked loop (PLL) and the high-power RF test at 2.0 K is done following a standard procedure used to test SRF cavities [14]. Temperature maps are taken as a function of $B_p$ to identify hotspot regions to be investigated. During high-power RF tests, a lead shield slides over the top plate of the test stand inserted into the cryostat and closure of the lid provides an interlock signal for safe operation of both the high-power laser and the high-power RF amplifier used to excite the cavity.

In order to perform LTLSM, a signal related to the PR(x, y) of the Nb surface is obtained in the following way: while the cavity is locked on the resonant frequency, with constant RF input power, the laser is switched on and directed at a particular location of the cavity flat plate. The laser power is pulsed from 0 to 100% with a TTL signal from a function generator (Agilent, model 33120A) at a frequency $f_M$. The laser beam diameter and average power can be adjusted as described in Sec. II.D. The same TTL signal is used as reference signal in a lock-in amplifier (Stanford Research Systems, model SR810). A fraction of the transmitted power signal, $P_t$, from the cavity is rectified through a crystal diode with matched load resistor, so that the output voltage is proportional to the amplitude of the RF signal. This signal is AC-coupled to the input of the lock-in amplifier and its amplitude is measured by the instrument and acquired by a LabVIEW program. A basic schematic of the measurement setup is shown in Fig. 9.

In steady state, the transmitted power from the cavity as a function of frequency $f$ and loaded Q, $Q_L$ is given by:

$$P_t(f, Q_L) = \frac{4 P_i}{Q_{ext1} Q_{ext2}} \frac{Q_L^2}{1 + 4 Q_L^2 (f/f_0 - 1)^2} = \frac{\alpha Q_L^2}{1 + 4 Q_L^2 (f/f_0 - 1)^2} \quad (1)$$

where $P_i$ is the incident power, $Q_{ext1}$ and $Q_{ext2}$ are the external Qs of the input and pick-up antenna, respectively and $f_0$ is the resonance frequency. The measurements described below will take place on resonance, $f = f_0$. A change in transmitted power due to a change in $Q_L$, measured at the resonance frequency, is given by:

$$\partial P_t(f, Q_L)\big|_{f=f_0} = 2 P_t(f_0, Q_L) \frac{\partial Q_L}{Q_L} = 2 P_t(f_0, Q_L) \left[1 - \left(\frac{Q_L}{Q_{ext1}} + \frac{Q_L}{Q_{ext2}}\right)\right] \frac{\partial Q_0}{Q_0} \quad (2)$$

If $P_t(f_0, Q_L)$ is the transmitted power signal $P_{t0}$ measured when the laser is off, the relative change of the $Q_0$ when switching on the laser is given by:

$$\frac{\partial Q_0}{Q_0} = -\frac{\partial P_c}{P_{c0}} \cong -\frac{1}{2} \frac{R_s(T_f, x, y) H^2(x, y) \pi r_L^2}{P_{c0}} \quad (3)$$

where $P_{c0}$ is the RF power dissipated in the cavity when the laser is off, $r_L$ is the radius of the laser beam on target (half of $D_L$), $R_s(T_f, x, y)$ and $H(x, y)$ are the surface resistance and surface magnetic field averaged over the beam area, respectively, and $T_f$ is the local temperature at the inner surface when the laser is on. In Eq. (3) we made the assumption that the surface resistance with the laser on is much greater than its value with the laser off, as the BCS-component of the surface resistance increases exponentially with temperature. Another assumption in Eq. (3) is that the local magnetic field is temperature independent. Inserting Eq. (3) into (2), the absolute change of $P_t$ when the laser is turned on at the $(x, y)$ location is given by:

$$\delta P_t(x, y, f_M) \cong \frac{P_{t0}}{P_{c0}} \left[1 - \left(\frac{Q_L}{Q_{ext1}} + \frac{Q_L}{Q_{ext2}}\right)\right] \cdot R_s(x, y, f_M) H^2(x, y) \pi r_L^2 \quad (4)$$

$Q_L$, $Q_{ext1}$, $Q_{ext2}$, $P_{t0}$ and $P_{c0}$ are all calculated from measurements of the incident, reflected and transmitted power along with the resonant frequency and the decay time of the transmitted power when the RF power is turned off. Since the cavity stored energy is also obtained from the measurement, the amplitude of the magnetic field on the cavity surface is also known from an RF simulation of the $TE_{011}$ mode in the cavity geometry. In Eq. (4) we included the dependence of $\delta P_t$ and $R_s$ on the laser intensity modulation frequency, which will affect $T_f$, and we assumed that the heating area is not affected by $f_M$, which is valid for low modulation frequencies.

When the laser is pulsed with a TTL signal, the voltage output from the crystal diode is a square-wave signal with a DC offset. The peak-to-peak value of this signal is proportional to $\delta P_t(x, y, f_M)$. The lock-in amplifier measures the rms value of the amplitude of the Fourier component of the signal with the same frequency as the reference signal and is given by:

$$\delta V_{rms}(x, y, f_M) = \frac{1}{\sqrt{2}} \frac{4}{\pi} \frac{\delta P_t(x, y, f_M)}{2\beta} \quad (5)$$

where $\beta$ is the power-to-voltage proportionality constant which is measured for the crystal diode with load resistor used in the experiments. Substituting (4) into (5), the local surface resistance is calculated from the signal measured by the lock-in amplifier:

$$R_s(x, y, f_M) \cong \frac{\pi}{\sqrt{2}} \frac{P_{c0} \delta V_{rms}(x, y, f_M)}{V_0 \left[1 - \left(\frac{Q_L}{Q_{ext1}} + \frac{Q_L}{Q_{ext2}}\right)\right] H^2(x,y) \pi r_L^2} \quad (6)$$

where $V_0 = \beta P_{t0}$ is the voltage output from the crystal detector when the laser is off. The surface resistance calculated from (6) is at the temperature of the region heated by the laser beam, which is higher than the He bath temperature.

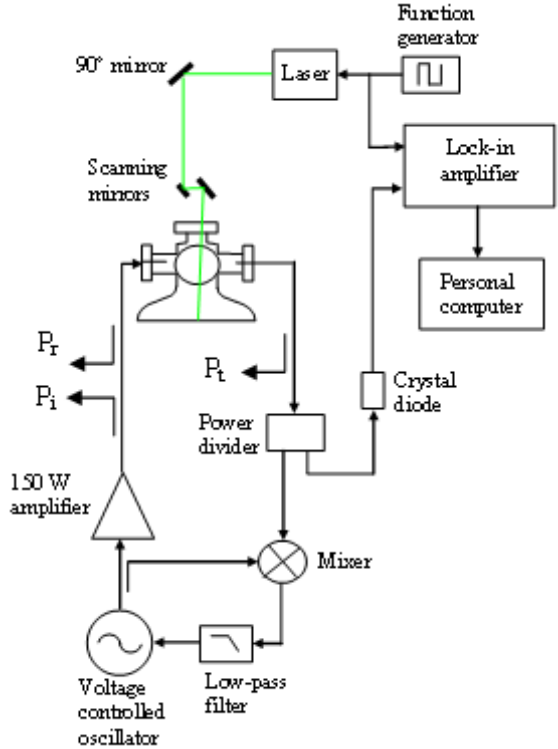

FIG. 9. (Color online) Basic schematic of the measurement setup.

An estimate of the fraction of the laser power absorbed into Nb at room temperature was done in the following way:

(a) the total attenuation of the optical components is the ratio of the power measured at the output of the viewport on the mirrors' chamber interface flange to the power measured at the output of the attenuator box

(b) the ratio of temperature rise on the outer Nb surface to the laser absorbed power, $\Delta T_{out}/P_{abs}$, was calculated from a thermal analysis of the Nb cavity plate at room temperature in air using a 2D finite element code (ANSYS [15]). A Gaussian laser beam profile was used in the calculation. The temperature peak on the outer cavity surface was measured with a thermocouple for a laser output power ($P_L$) of 10 W and a beam diameter of 0.87 mm and $P_{abs}$ was then obtained from $\Delta T_{out}/P_{abs}$. The result indicates that ~ 21% of the laser power from the attenuator box is absorbed in the Nb cavity plate. Since the attenuation due to the optical components was measured, the absorptivity of the inner surface of the Nb plate at 532 nm was calculated to be ~34%, consistent with results in Ref. [11]. No major change in absorptivity between 295 K and 4.2 K was also reported [11].

The efficiency of the temperature sensors measuring the temperature of the outer surface of the cavity plate immersed in liquid He at 2.0 K was estimated as follows:

(a) a numerical thermal analysis of the Nb cavity plate (10 cm radius and 2.36 mm thick, as measured on the cavity after fabrication and chemical etching) was done with ANSYS using temperature dependent thermal conductivity and Kapitza conductance [16] to calculate the temperature distributions on the inner and outer surfaces of the plate, heated with a Gaussian beam at its center, as a function of beam diameter and power. The dependence of the maximum temperature rise on the outer surface, $\Delta T_{out}$, on laser power and beam diameter on target can be fitted well with a straight line:

$\Delta T_{out}^{calc}(K) = -0.096\ D_L(mm) + 2.901\ K$

$P_{abs} = 2$ W, $0.87$ mm $\leq D_L \leq 3.02$ mm

$\Delta T_{out}^{calc}(K) = 0.96\ P_{abs}(W) + 0.886\ K$

$D_L = 0.87$ mm, $1$ W $\leq P_{abs} \leq 2$ W.

(b) The peak temperature rise on the outer surface was measured with a temperature sensor in contact with the cavity plate immersed in 2.0 K He bath for different laser power levels and beam diameters. The average thermometer efficiency, defined as the ratio $\Delta T_{out}^{meas}/\Delta T_{out}^{calc}$ was ~20%. The dependence of $\Delta T_{out}^{meas}$ on the beam diameter was measured to be as follows:

$\Delta T_{out}^{meas}(K) = -0.097\ D_L(mm) + 0.675\ K$

$P_{abs} = 2$ W, $0.87$ mm $\leq D_L \leq 3.02$ mm

The slope is in agreement with the value in Eq. (7) and the ratio of the intercepts (0.675/2.901 = 23%) is in agreement with the thermometer efficiency obtained from the power dependence of $\Delta T_{out}^{meas}$.

The thermometers' efficiency is limited by factors such as insufficient contact force, cracks in the insulation and the quality of the thermal bond between the thermometer and the cavity surface.

Figure 10 shows a plot of the equilibrium peak temperature at the inner ($T_{in}$) and outer ($T_{out}$) surfaces of the cavity plate due to heating from a 0.87 mm diameter continuous-wave laser beam, calculated with ANSYS. The outer surface is in contact with the He bath at 2.0 K. The figure shows that an absorbed power of ~100 mW is already sufficient to raise the inner temperature by almost 3 K.

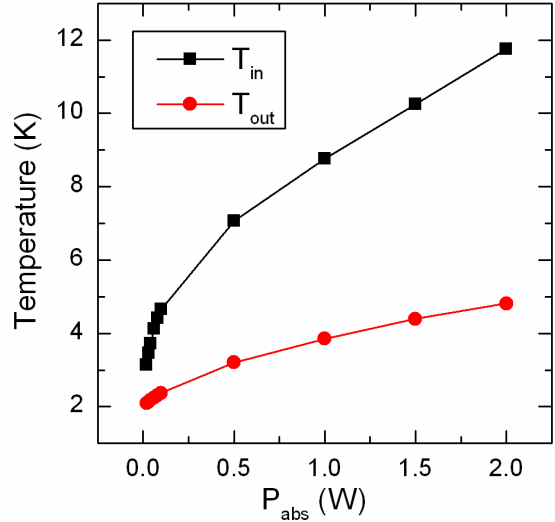

FIG. 10. (Color online) Peak temperature at the inner and outer surfaces as a function of the absorbed laser power (beam diameter = 0.87 mm, continuous-wave) calculated with ANSYS. The outer surface is in contact with the He bath at 2.0 K.

## IV. EXPERIMENTAL RESULTS

Figures 11(a)-(b) show $\delta V_{rms}$, normalized to the maximum value at 1 Hz, measured by the lock-in amplifier as a function of modulation frequency and absorbed power. The amplitude of $\delta V_{rms}$ increases with increasing laser power and lower modulation frequency. The data indicating that $\delta V_{rms}(f_M)$ decreases more rapidly at higher laser power suggests an increased response time to reach thermal equilibrium. This can be due to a reduced heat transfer at the Nb-He interface, as the numerical simulation indicates reaching a peak outer surface temperature of the plate corresponding to the so-called lambda-point (2.17 K) for an absorbed laser power of ~60 mW.

Figures 12(a)-(b) show the result of a horizontal line-scan, $\delta V_{rms}(x, 1\ cm)$, at 2.0 K, $B_p = 13$ mT, $f_M = 1$ Hz, $P_{abs} = 0.88$ W, which is consistent with the calculated surface magnetic field distribution in the TE$_{011}$ mode, as shown. The values of the parameters in Eq. (6) are: $V_0 = 17$ mV, $P_{c0} = 229$ mW, $Q_L = 1.22 \times 10^9$, $Q_{ext1} = 2.45 \times 10^9$, $Q_{ext2} = 5.21 \times 10^{10}$, $D_L = 0.87$ mm. Figure 12(b) shows higher surface resistance within ~10 mm from the center of the plate. The scan step was 1 mm and time constant setting in the lock-in amplifier was 1 s, 12 dB/oct.

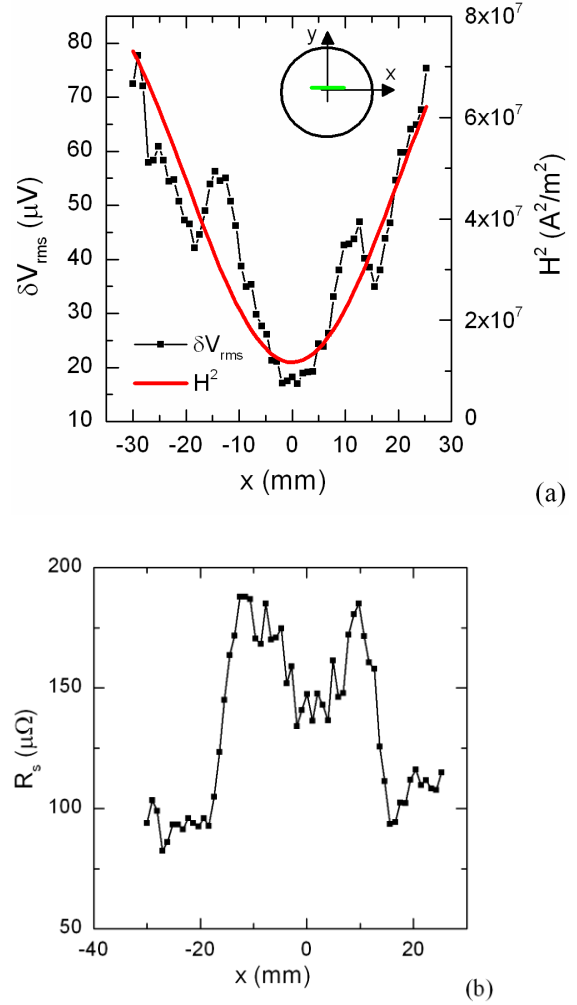

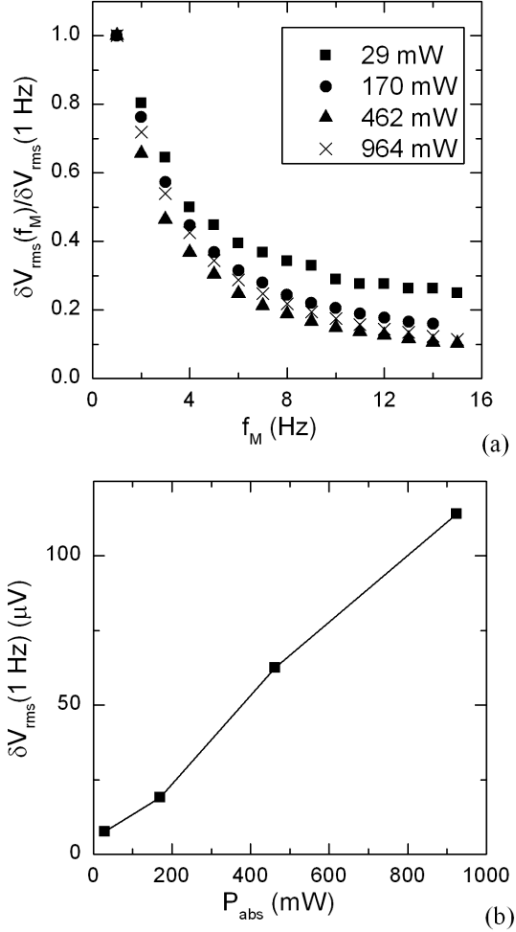

FIG. 12. (Color online) (a) Signal measured with the lock-in amplifier while scanning the laser ($P_{abs} = 0.88$ W, $D_L = 0.87$ mm, $f_M = 1$ Hz) along the horizontal direction and squared magnetic field in the TE$_{011}$ mode at the same location. The location of the line-scan on the cavity plate is shown in the inset. (b) The surface resistance calculated with Eq. (6).

FIG. 11. (a) Normalized amplitude of the signal from the lock-in amplifier, $\delta V_{rms}$, as a function of the laser modulation frequency, for different values of $P_{abs}$. (b) $\delta V_{rms}$ as a function of $P_{abs}$ at 1 Hz. The laser beam diameter is 0.87 mm.

The estimated temperature of the inner surface for $P_{abs} = 0.88$ W, $D_L = 0.87$ mm is 8.3 K and the surface resistance at 3.3 GHz calculated using a numerical code [17] based on the Bardeen-Cooper-Schrieffer (BCS) theory of superconductivity ($R_{BCS}$) is ~98 μΩ, in good agreement with the values shown in Fig. 12(b) for $|x| > 12$ mm. Nevertheless, the surface resistance was not uniform along the scan direction and was almost a factor of two higher than $R_{BCS}$ close to the center.

Figure 13 shows the temperature map of the cavity flat plate at a He bath temperature of 2.0 K and $B_p$ = 13 mT. 2D scans of the regions highlighted with the rectangular box in the figure were done with different laser parameters such as power and modulation frequency. Images of $R_s(x, y)$ are shown in Figs. 14-15 for different values of $P_{abs}$ and $f_M$. In all cases, the x-y scan step size was 1 mm and $D_L$ = 0.87 mm. The lock-in amplifier time constant was reduced to 0.3 s when the modulation frequency was 10 Hz or 100 Hz to reduce the data acquisition time. The temperature of the plate's inner surface estimated by comparing the $R_s$-values shown in Figs. 14(b)-(d) with a numerical calculation of the BCS surface resistance, are ~8.5 K for $f_M$ = 1 Hz, ~6.5 K for $f_M$ = 10 Hz and ~4 K for $f_M$ = 100 Hz.

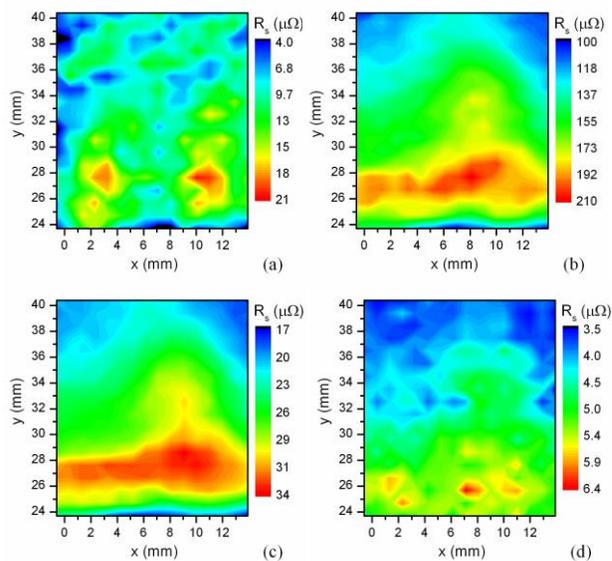

FIG. 14. (Color online) Surface resistance maps at $B_p$ = 13 mT on a 16 mm × 18 mm area (region 1 in Fig. 13) of the cavity flat plate obtained for different laser parameters: $P_{abs}$ = 50 mW, $f_M$ = 1 Hz (a), $P_{abs}$ = 0.92 W, $f_M$ = 1 Hz (b), $P_{abs}$ = 0.92 W, $f_M$ = 10 Hz (c), $P_{abs}$ = 0.92 W, $f_M$ = 100 Hz (d).

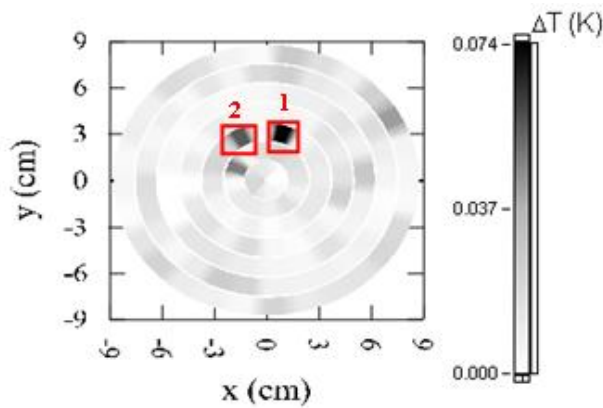

FIG. 13. (Color online) Temperature map of the cavity flat plate at $B_p$ = 13 mT and He bath temperature of 2.0 K. The rectangular boxes show two hotspot regions which have been analyzed with LTLSM.

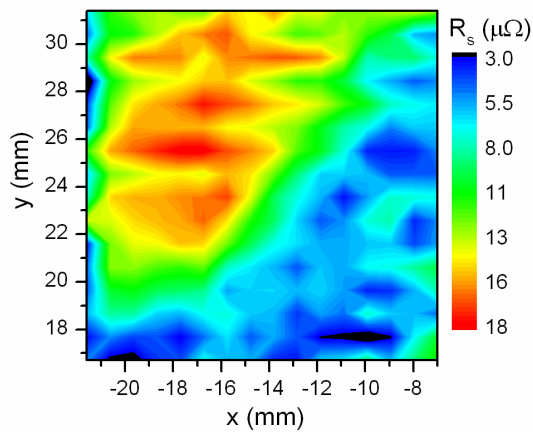

FIG. 15. (Color online) Surface resistance map $B_p$ = 13 mT on a 16 mm × 16 mm area (region 2 in Fig. 13) of the cavity flat plate obtained with laser parameters $P_{abs}$ = 50 mW, $f_M$ = 1 Hz.

The temperature map at 2.0 K, $B_p$ = 13 mT from another cavity RF test is shown in Fig. 16. Figure 17(a) shows the surface resistance map obtained using LTLSM over the area indicated by the red box in Fig. 16. Figure 17(b) shows the surface resistance map for a fraction of the area shown in Fig. 17(a) which was scanned with a smaller x-y scan step (0.5 mm compared to 1 mm). The laser parameters were $P_{abs} = 0.9$ W and $f_M = 1$ Hz, $D_L = 0.87$ mm. Figure 17(c) shows the surface resistance map for a fraction of the area shown in Fig. 17(a) and scanned with lower laser power ($P_{abs} = 0.46$ W) and higher RF field ($B_p = 50$ mT) than in Figs. 17(a)-(b).

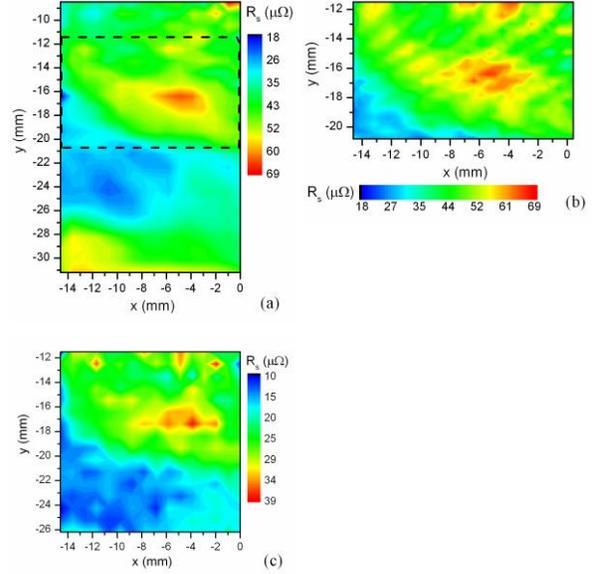

FIG. 17. (Color online) Surface resistance map at $B_p = 13$ mT and He bath temperature of 2.0 K over a 15 mm × 23 mm area (highlighted in Fig. 16) obtained with laser parameters $P_{abs} = 0.9$ W, $f_M = 1$ Hz, $D_L = 0.87$ mm and x-y scan step of 1 mm (a). The region in the dashed box in (a) has been scanned with higher x-y resolution (0.5 mm), shown in (b). A fraction of the area shown in (a) was also scanned with lower laser power ($P_{abs} = 0.46$ W), x-y scan step of 1 mm but higher RF field ($B_p = 50$ mT) (c).

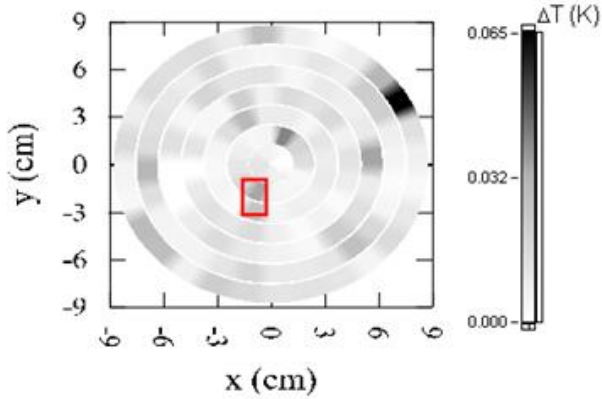

FIG. 16. (Color online) Temperature map of the cavity flat plate at $B_p = 13$ mT and He bath temperature of 2.0 K during a new RF test. The rectangular boxes shows a hotspot region which has been analyzed with LTLSM.

## V. DISCUSSION

The results shown in the previous section demonstrate the feasibility of the LTLSM technique to study RF losses in "real size" SRF cavities. Different parameter values were tried during the measurements to improve the signal-to-noise ratio (SNR). As a result, it appears that there has to be a compromise between high SNR, which can be obtained with higher laser power and low modulation frequency, and faster data acquisition and/or low temperature rise at the inner surface. SNR was calculated as the average from the ratio of the voltage from the lock-in amplifier measured while scanning the laser with RF on, divided by the voltage with RF off:

$$SNR = 10 \log\left[\left(\delta V_{rms}^{RFon} / \delta V_{rms}^{RFoff}\right)^2\right]. \quad (10)$$

The SNR value for $P_{abs} = 50$ mW, $f_M = 1$ Hz, $B_p = 13$ mT was ~10 dB, increasing to ~30 dB for $P_{abs} = 0.9$ W.

The sensitivity of this technique can be calculated from Eq. (6) as the ratio of the additional RF power density dissipated when the laser is on, $p_{RF} = 1/2 R_s H^2$, divided by the voltage measured with the lock-in amplifier, $\delta V_{rms}$. Using $D_L = 0.87$ mm and typical values of the other parameters in Eq. (6), we obtain $p_{RF}/\delta V_{rms} \sim 84$ µW/(mm² µV) for $B_p = 16$ mT. The $R_s$-sensitivity depends on the amplitude of the magnetic field at the location where $R_s$ is measured. At the location of peak surface magnetic field, the $R_s$-sensitivity is ~ 1 µΩ/µV. The noise in the measurement of $\delta V_{rms}$ is about 0.8 µV, corresponding to a resolution in the $R_s$ measurement of ~ 0.8 µΩ, at the peak surface magnetic field location. At the same location, increased resolution and sensitivity could be obtained by improving the measurement of the transmitted signal from the cavity, for example by replacing the diode with a device with greater dynamic range and lower noise floor. In addition, a smaller laser spot size would also increase the $R_s$-sensitivity, besides improving the spatial resolution.

The power of the laser beam reflected off the cavity flat plate will be dissipated into the He bath through multiple reflections inside the cavity volume when the incoming beam contacts the cavity plate at a distance greater than about 9 mm from the center. The power density of the beam will decrease rapidly after each reflection, because of the roughness of the surface and the absorption by the Nb walls, without causing any significant overheating of the Nb surface. If the beam entering the cavity contacts the cavity plate at a radial distance lower than about 9 mm, it will be reflected through the window into the mirrors' chamber and it will dissipate its power onto the walls of the long vertical tube inside the cryostat.

A comparison of the LTLSM technique used in this study with the ones used to investigate the superconducting properties of HTS samples is discussed in what follows. Firstly, we estimate parameters such as the laser penetration depth, the thermal diffusion length and the thermal response time for our system, which will be useful when comparing different experimental setups.

The laser penetration depth in the Nb surface, $d_L$, is given by:

$$d_L = \frac{\lambda}{4\pi k}, \quad (11)$$

where $\lambda$ is the laser wavelength and $k$ is the imaginary part of the refractive index. At 532 nm, $k = 2.88$ [18], resulting in $d_L = 15$ nm. The thermal diffusion length, $d_T$, can be calculated from the 2D heat diffusion equation as [19]:

$$d_T = \sqrt{\frac{\kappa d}{h_K}}, \quad (12)$$

where $\kappa$ is the thermal conductivity of Nb at 2 K, of the order of 10 W/(m K), $d$ is the wall thickness, which was measured to be 2.4 mm and $h_K$ is the Kapitza conductance at the Nb-He interface, of the order of about 5 kW/(m² K). The thermal diffusion length estimated from Eq. (12) at 2 K is ~2.2 mm, of the order of the wall thickness. Similarly, the thermal response time $\tau$ of the cavity plate for reaching thermal equilibrium after laser heating can be estimated as [20]:

$$\tau = \frac{C d}{h_K}, \quad (13)$$

where $C$ is the specific heat per unit volume at an intermediate temperature between those of the inner and outer surfaces, ~3 K, of the order of 750 J/(m³ K) [21], resulting in $\tau \sim 0.36$ ms. At high laser power ($P_{abs} > 60$ mW), the temperature on the outer surface of the cavity plate increases above the lambda-point. The Kapitza conductance of He I is reduced by about two orders of magnitude compared $h_K$ in He II, increasing τ to ~30 ms. Even longer response time can be expected due to more complex convection mechanisms, when transitions to nucleate or film boiling occur in liquid He for progressively higher laser power density. The values of $d_T$ and τ at 4.2 K in thin film HTS studied with LTLSM range between 1-10 μm and 0.1-10 μs [22]. In the limit of high modulation frequency ($f_M τ \gg 1$), the thermal diffusion length is reduced by a factor $1/\sqrt{\omega_M \tau}$ ($\omega_M = 2\pi f_M$), therefore modulation frequencies greater than about 100 kHz are often used to improve the spatial resolution by reducing $d_T$ to few micrometers. In our experiments, we found that the amplitude of $\delta V_{rms}$ is of the order of few μV for $f_M \sim 100$ Hz and therefore $d_T$ cannot be reduced without loss of sensitivity in the measurement. The surface resistance maps in Figs. 14, 15 and 17 show features of the size of few millimeters, of the order of $d_T$. As a general comment, it can be noted that all major lengths in the LTLSM setup for the SRF cavity increase by almost three orders of magnitude compared to the setup for HTS samples [6]: the spatial resolution increases from a few micrometers to 1-2 millimeters, the sample size increases from few millimeters to tens of centimeters, the size of the apparatus increases from tens of centimeters to few meters.

The RF imaging mode commonly used in LTLSM of thin film HTS superconductors allows obtaining maps of the local current density, as the perturbation due to the laser mostly changes the resonant frequency but not the quality factor. In the case of an SRF Nb cavity, which has $Q_0$ values of the order of $10^9$, compared to $\sim 10^4$ for HTS strip line resonators, the effect of the laser heating is mostly a change of the RF dissipated power, which allows obtaining maps of the surface resistance. Because the stored energy in the cavity is much greater than that of a strip line resonator, the relative change of the resonant frequency due to a spot of 1 mm radius with temperature ~8 K is less than $10^{-9}$ (less than 3 Hz at 3.3 GHz). This change is much smaller than frequency variations due, for example, to the sensitivity to He bath pressure variations, which was measured to be ~ 3.4 kHz/mbar, and to the Lorentz force detuning coefficient, which was measured to be ~ -3.3 kHz/(mT)$^2$. Furthermore, the cavity is constantly locked to the resonant frequency by the PLL during the measurements.

Several possibilities can explain the non-uniform surface resistance, observed with this technique and, with lower spatial resolution, by thermometry. Those include grain boundaries or other crystal defects, impurities such as oxygen or hydrogen, losses in the surface oxide layer and/or pinned vortices. As will be described in detail in a future publication, the experimental results showed evidence for changes in the intensity and location of the hotspots after laser scanning or RF cycling, which suggest that pinned vortices are one good candidate to explain the origin of "anomalous" RF losses [23].

# VI. CONCLUSION

LTLSM is a well known technique to study loss mechanisms in HTS samples. In this contribution we describe in detail the experimental apparatus and results of applying the basics of this technique to obtain, for the first time, 2D maps of the surface resistance of regions of the inner surface of a "real-size" SRF cavity. The application of this technique to SRF cavities improves the spatial resolution for locating RF hotspots from ~1-2 cm to ~1-2 mm. In order to maintain a good SNR, the laser modulation frequency has to be of the order of a few Hz. Samples cut from the SRF cavity plate will be examined by optical microscopy and surface analytical techniques to further investigate the origin of the anomalous RF losses.

# ACKNOWLEDGMENTS

The authors would like to acknowledge P. Kushnick and B. Clemens from Jefferson Lab for cryogenic support and electron-beam welding of the Nb cavity and Prof. A. Gurevich from Old Dominion University for many valuable discussions. This manuscript has been authored by Jefferson Science Associates, LLC under U.S. DOE Contract No. DE-PS02-09ER09-05. Additional support for this work was provided the U.S. Government Presidential Early Career Award for Scientists and Engineers. The U.S. Government retains a non-exclusive, paid-up, irrevocable, world-wide license to publish or reproduce this manuscript for U.S. Government purposes. S.M.A. acknowledges support from U.S. DOE (DESC 0004950) and the UMD/ONR Applied Electromagnetics Center, task D10 (N000140911190), and the Maryland Center for Nanophysics and Advanced Materials.